\documentclass[reprint,superscriptaddress]{revtex4-1}
\usepackage{graphicx,subfigure,epsfig,appendix,bm,tabularx}
\usepackage{booktabs,multirow,mathrsfs,amssymb,amsmath,amsthm}
\usepackage{framed,listings} 
\usepackage[colorlinks=true,linkcolor=blue,citecolor=blue,urlcolor=blue]{hyperref}
\usepackage[noend]{algpseudocode}\usepackage[ruled]{algorithm2e}
\newcommand{\bra}[1]{\langle #1|}\newcommand{\ket}[1]{|#1\rangle}
\newcommand{\inner}[2]{\langle #1|#2 \rangle}
 
\usepackage[usenames]{color}
\usepackage{qcircuit}
\definecolor{red}{rgb}{1,0,0}\definecolor{blue}{rgb}{0,0,1}
\begin{document}


\title{Quantum Resonant Dimensionality Reduction and Its Application in Quantum Machine Learning}

\author{Fan Yang}\altaffiliation{These authors contributed equally to this work.}
\affiliation{Beijing Academy of Quantum Information Sciences, Beijing 100193, China}
\affiliation{State Key Laboratory of Low-Dimensional Quantum Physics and Department of Physics, Tsinghua University, Beijing 100084, China}
\author{Furong Wang}\altaffiliation{These authors contributed equally to this work.}%
\affiliation{China North Vehcle Research Institute, Beijing 100072, China}

\author{Xusheng Xu}
\affiliation{State Key Laboratory of Low-Dimensional Quantum Physics and Department of Physics, Tsinghua University, Beijing 100084, China}

\author{Pao Gao}
\affiliation{Beijing Academy of Quantum Information Sciences, Beijing 100193, China}

\author{Tao Xin}
\affiliation{Shenzhen Institute for Quantum Science and Engineering, Southern University of Science and Technology, Shenzhen 518055, China}

\author{ShiJie Wei}\email{weisj@baqis.ac.cn}
\affiliation{Beijing Academy of Quantum Information Sciences, Beijing 100193, China}

\author{Guilu Long}\email{gllong@tsinghua.edu.cn}
\affiliation{State Key Laboratory of Low-Dimensional Quantum Physics and Department of Physics, Tsinghua University, Beijing 100084, China}
\affiliation{Tsinghua National Laboratory for Information Science and Technology, Beijing 100084, People’s Republic of China}
\affiliation{Collaborative Innovation Center of Quantum Matter, Beijing 100084, China}


\begin{abstract}

Quantum computing is a promising candidate for accelerating machine learning tasks. Limited by the control accuracy of current quantum hardware, reducing the consumption of quantum resources is the key to achieving quantum advantage. Here, we propose a quantum resonant dimension reduction (QRDR) algorithm based on the quantum resonant transition to reduce the  dimension of input data and accelerate the quantum machine learning algorithms. After QRDR, the dimension of input data  $N$ can  be  reduced into desired scale $R$, and the effective information of the original data will be preserved correspondingly, which will reduce the computational complexity of subsequent quantum machine learning algorithms or quantum storage. QRDR operates with polylogarithmic time complexity and reduces the error dependency from the order of $1/\epsilon^3$ to the order of $1/\epsilon$, compared to existing algorithms. We demonstrate the performance of our algorithm combining with two types of quantum classifiers, quantum support vector machines and quantum convolutional neural networks, for classifying underwater detection targets and quantum many-body phase respectively. The simulation results indicate that reduced data improved the processing efficiency and accuracy  following the application of QRDR. As quantum machine learning continues to advance, our algorithm has the potential to be utilized in a variety of computing fields.

\end{abstract}

\maketitle
\section{introduction}

Machine learning has proven to be a remarkably powerful tool with broad practical applications across various fields of science and engineering, including finance \cite{GOODELL2021100577, GHODDUSI2019709}, medical science \cite{doi:10.1148/rg.2017160130, doi:10.1161/CIRCULATIONAHA.115.001593}, and the simulation of classical and complex quantum systems. It has shown notable success in compressing high-dimensional data, where near-term quantum devices may offer significant speed enhancements \cite{gujju2023quantum}. Given the potential  quantum advantage of quantum computing, numerous quantum machine learning (QML) algorithms have been proposed \cite{Dunjko_2018}, such as quantum recommendation systems (QRS) \cite{kerenidis2016quantum}, quantum support vector machines (QSVM) \cite{PhysRevLett.113.130503}, quantum principal component analysis (QPCA) \cite{lloyd2014quantum}, and quantum neural networks \cite{abbas2021power, beer2020training}, among others. 

In machine learning, dimensionality reduction (DR) is a valuable technique for refining information and significantly decreasing data processing time. DR involves compressing high-dimensional datasets into lower-dimensional representations while preserving key information from the original dataset.
 Principle component analysis (PCA) DR is the most representative example of linear DR. It works by projecting the original data onto the subspace of the covariance matrix with larger singular values while disregarding components with smaller values. These smaller principal components often represent noise, so dimensionality reduction can enhance the accuracy of subsequent data processing. By applying kernel methods, kernel principal component analysis (KPCA) extends the concept of PCA to enable nonlinear dimensionality reduction.

 In recent years, several quantum dimensionality reduction (QDR) methods have been proposed.  Here we focus on an important type, the  PCA-based QDR algorithms,  which can efficiently remove redundant information by identifying orthogonal principal components, reducing computational complexity, and enhancing model performance, especially in high-dimensional datasets. Among PCA-based QDR algorithms, the most renowned one is quantum principal component analysis (QPCA) \cite{lloyd2014quantum}, which uses density matrices to obtain their eigenvalues and eigenstates.  In QPCA, quantum amplitude estimation and quantum phase estimation (QPE) are also necessary for preparing data vectors after DR, but they demand significant quantum resources. Following QPCA, two approaches can achieve QDR, one is QPE-based QDR \cite{Li_2021, Yu.2019} and the other is variational quantum algorithm (VQA)-based QDR \cite{PhysRevA.101.032323,PhysRevA.101.032323}.QPE-based algorithms have reduced time complexity compared to QPCA. They still face challenges related to the required number of ancilla qubits and evolution time, posing difficulties for current state-of-the-art quantum computers. In contrast, VQA-based algorithms are generally seen as friendly for noisy intermediate-scale quantum (NISQ) devices, but its complexity becomes increasingly challenging to analyze as the number of qubits grows.

 There are also many other types of quantum dimension reduction algorithms, like based on autoencoders \cite{Romero_2017, huang2020realization}, A-optimal projection \cite{PhysRevA.99.032311, PhysRevA.102.052402}, linear discriminant analysis \cite{YU2023128554, cong2016quantum} and locally linear embedding \cite{he2020quantum, 8807145}. These algorithms are based on different data mining principles and are used in different data sets and different demand scenarios. We don't go to details here.

 To address these challenges, we introduce a novel quantum dimensionality reduction  algorithm called quantum resonant dimensionality reduction (QRDR), based on quantum resonant transitions (QRT) \cite{PhysRevLett.122.090504, RQPCA_Li}.  By employing QRT instead of QPE, the number of ancilla qubits is reduced from $O(1/\epsilon+\log R)$ to $O(\log R)$ in QRDR that is independent of the accuracy $\epsilon$, offering an advantage over other quantum DR algorithms such as QPE-based methods \cite{Li_2021, Yu.2019}.  In contrast to VQA-DR, our QRDR presents an exact time complexity. Comnpared to QPE-based DR algorithm and QPCA algorithm, the time complexity of QRDR is reduced from the order $1/\epsilon^3$ to the order $1/\epsilon$. We simulate the QRDR algorithm numerically and evaluate its performance using quantum support vector machines and quantum convolutional neural networks (QCNN). The results demonstrate that our QRDR effectively reduces the dimension of quantum data, enabling various quantum machine learning algorithms to process the data more efficiently post-reduction.

 The remainder of this paper is organized as follows: Sec. \ref{framwork} provides the framework of the algorithm and details the quantum resources consumed. Sec. \ref{numer} presents the simulation results of applying QRDR in  quantum support vector machines and quantum convolutional neural networks. In Sec. \ref{dis}, we compare our algorithm with existing  approaches and conclude with a summary of our findings.

\section{The framwork of the algorithm }\label{framwork}
\subsection{Principal component analysis dimension reduction and its quantum expression}
Principal component analysis  for dimensionality reduction relies on linear transformations involving the covariance matrix (or kernel matrix). In this paper, we focus on the basic PCA method, leaving out advanced techniques such as kernel methods and centering, which are detailed in references \cite{Li_2021, Yu.2019}.

 The original data before  DR can be represented as a matrix $X \in \mathbb{R}^{M \times N}$, where the $i$th row of $X$ corresponds to the $i$th sample, $\bm{x_i} = [x_i^1, x_i^2, \ldots, x_i^N]$, containing $N$ features. The covariance matrix of this data, denoted as $A \in \mathbb{R}^{N \times N} = X^TX$. The data undergoes singular value decomposition (SVD) as follows
\begin{equation}
	X=\sum_{k=1}^{M}\sigma_k\bm{u_k}^T \bm{v_k},
\end{equation}
where $\bm{u_k}$ and $\bm{v_k}$ are singular vectors corresponding to singular value $\sigma_k$.  It is usually assumed that the singular values are ordered from largest to smallest. The matrix $A$ has eigenvalue decomposition
\begin{equation}
	A=\sum_{k=1}^{M}\sigma_k^2 \bm{v_k}^T \bm{v_k}.
\end{equation}
By definition, $A$ is the Hermitian matrix and the $k$-th eigenvalue is $\lambda_{k}=\sigma_k^2$ with eigenvector $\bm{v_k}$. When the dimension is reduced from $N$ to $R$, the data matrix $Z \in \mathbb{R}^{M \times R}$ after dimensionality reduction can be represented by the formula $Z=XV_R^T$. The matrix $V_R$ is a projection matrix composed of the first $R$ singular vectors, with each row $j$ of $V_R$ corresponding to the $j$th singular vector $\bm{v_j}$. Consequently, the value of the $j$-th feature for $z_i$ is given by $z_i^j = \bm{x_i} \cdot \bm{v_j}^T$.

For the  QDR, the input is a quantum state that encodes the information from the original data
\begin{equation}\label{Eq:inpustate}
	\ket{X}=\sum_{i=1}^{M}\sum_{j=1}^{N}x_i^j\ket{j}\ket{i}=\sum_{i=1}^{M}\|x_i\|\ket{x_i}\ket{i},
\end{equation}
where $\ket{i}$ and $\ket{j}$ represent quantum computational basis states, and $\ket{x_i} = \frac{1}{|x_i|} \sum_{j=1}^{N} x_i^j \ket{j}$ with the normalized coefficient $|x_i| = \sqrt{\sum_{j=1}^{N} |x_i^j|^2}$. The output state of  QDR is given by
\begin{equation}\label{eq:aim}
	\ket{Z}=\sum_{i=1}^{M}\sum_{j=1}^{R}z_i^j\ket{j}\ket{i}=\sum_{i=1}^{M}\|z_i\|\ket{z_i}\ket{i}.
\end{equation}
The goal of QDR is to prepare the state $\ket{Z}$ as described in Eq.(\ref{eq:aim}). In this section, the overall normalization coefficients for the quantum states are omitted.

\subsection{Quantum Resonant Dimensionality Reduction Algorithm}
We use quantum resonant transitions (QRT) to realize the dimensionality reduction process, which includes designing the Hamiltonian and implementing the time evolution operator. Here we take one of the $M$ samples,  $N$ dimensional vector $\bm{x}$,  as the original data.
Our algorithm utilizes one probe qubit to find the eigenstate, $n = \lceil \log_2 N \rceil$ qubits to store the state corresponding to the original data, and $r = \lceil \log_2 R \rceil$ qubits to store the state after dimensionality reduction.  The Hamiltonian is defined as follows
\begin{equation}
	\label{Eq:Hamiltonian}
	\begin{aligned}
		\mathcal{H}=&\ket{0}\bra{0}\otimes(\ket{0}\bra{0}^{\otimes r}-I_2^{\otimes r})\otimes I_2^{\otimes n}+\ket{1}\bra{1}\otimes H_p\\
		&+\frac{c\pi}{2}\sigma_y\otimes B\otimes I_2^{\otimes n}.
	\end{aligned}
\end{equation}
Here $c$ represents the resonant parameter, $\sigma_y$ is the Pauli matrix, and $B = \sqrt{2^r}H_d^{\otimes r}$, where $H_d = \frac{1}{\sqrt{2}}[1,1;1,-1]$ is the Hadamard matrix. The three terms in Equation (\ref{Eq:Hamiltonian}) correspond to the Hamiltonian of the probe qubit, the problem Hamiltonian, and their interaction, respectively. Specifically, the second term, $H_p$, can be expressed as follows
\begin{equation}
	H_p=H_\lambda\otimes I_2^{\otimes n}+I_2^{\otimes r}\otimes A,
\end{equation}
where $H_\lambda$ is a diagonal matrix that encodes the $R$ eigenvalues $\lambda_k= \sigma_k^2$ of the covariance matrix $A$. The diagonal elements of $H_\lambda$ are given by $h_k = -\lambda_k $.
Given the original data $\bm{x}$, the input state  $\ket{\psi}_{in}$ is represented as $\ket{0}\ket{0}^{\otimes r}\ket{x}$. This typically requires specialized quantum devices such as quantum random access memory (QRAM) \cite{PhysRevLett.100.160501} to implement. The state can also be expanded in terms of the eigenvectors of $A$ as
\begin{equation}\label{Eq:input}
	\ket{\psi}_{in}=\ket{0}\ket{0}\ket{x}=\ket{0}\ket{0}^{\otimes r}\sum_{k=1}^{N}z^k\ket{v_k},
\end{equation}
where $z^k = \langle v_k | x\rangle$.
Setting $t = 1/c$, the state after the application of the Hamiltonian evolution operator $e^{-i\mathcal{H}t}$ is
\begin{equation}\label{Eq:evo}
	\begin{aligned}
		e^{-i\mathcal{H}t}\ket{\psi}_{in}&=\ket{1}\sum_{k=1}^{R}z^k\ket{k}\ket{v_k}\\
		&+\ket{0}\sum_{k'=R+1}^{N}\ket{0}^{\otimes r}z^{k'}\ket{v_{k'}}.
	\end{aligned}
\end{equation}
It is assumed that $c$ is much smaller than the minimum difference between eigenvalues of $H$, allowing us to neglect the effects of off-resonance. A detailed analysis of the dynamics, errors, and time complexity will be provided in the next section. 

To obtain the low-dimensional state, the probe qubit must be  collapsed   to state  $\ket{1}$ and the second and third registers disentangled. It means that we need to perform additional operations to make the quantum state evolve from $\ket{k}\ket{v_k}$ to $\ket{k}\ket{0}$, for each $k$. We employ the method proposed in Ref. \cite{Yu.2019}, which utilizes a reference state $\ket{\psi}$ and its corresponding inverse unitary operator $U_\psi^{-1}\ket{\psi}=\ket{0}^{\otimes n}$ to achieve this step.The projection $\phi_k$ of the reference state and each eigenstate $v_k$ needs to be measured. Then, apply the inverse unitary operator and controlled rotations to the ancilla qubits and work qubits for different wave vectors $k$. The final state is obtained in the second register is
\begin{equation}
	\ket{\psi}_{out}=\ket{1}\sum_{k=1}^{R}z^k\ket{k} \ket{0}^{\otimes n}.
\end{equation}

According to the above derivation, we take $\ket{X}$ that defined in Eq. \ref{Eq:inpustate} as original data state, the input state can be expressed as 
\begin{equation}
	\ket{\Psi}_{in}=\ket{0}\ket{0}^{\otimes r}\ket{X}=\ket{0}\ket{0}^{\otimes r}\sum_{i=1}^{M}\|x_i\|\ket{x_i}\ket{i}.
\end{equation}
The evolution operator does not affect the serial number state $\ket{i}$, so the state after applying the operator is
\begin{equation}
	\ket{1}\sum_{i=1}^{M}\sum_{k=1}^{R}z_i^k\ket{k}\ket{v_k}\ket{i}+
	\ket{0}\ket{0}^{\otimes r}\sum_{i=1}^{M}\sum_{k'=R+1}^{N}z_i^{k'}\ket{v_{k'}}\ket{i},
\end{equation}
where $z_i^k = \langle v_k | x_i\rangle$. 
After measuring the probe qubit and disentangling the second and third registers, the final state is consistent with Eq. (\ref{eq:aim})
\begin{equation}
	\ket{\Psi}_{out}=\ket{1}\sum_{i=1}^{M}\|z_i\|\ket{z_i}\ket{i}\ket{0}^{\otimes n}=\ket{1}\ket{Z}\ket{0}^{\otimes n},
\end{equation}
which is the output state of the algorithm. The QRDR process can be summarized in Alg.\ref{alg:QRDR}
\begin{algorithm}
	\label{alg:QRDR}
	\caption{Quantum Resonant Dimensionality Reduction}
	\LinesNumbered
	\textbf{Inputs:} Dataset state $\ket{X}$, covariance matrix $A$ and its first $R$ eigenvalues $\lambda_{k},k\in[1,R]$.\\
	\textit{step 1:} Prepare the initial state $\ket{0}\ket{0}^{\otimes r}\ket{X}$.\\
	\textit{step 2:} Set appropriate resonant parameter $c$.\\
	\textit{step 3:} Simulate the Hamiltonian $\mathcal{H}$ for a duration time of $t = 1/c$.\\
	\textit{step 4:} Measure the probe qubit and proceed if the measurement result is $\ket{1}$.\\
	\textit{step 5:} Disentangle the second and third registers using $U_\psi ^{-1}\ket{\psi}$.\\
	\textbf{Outputs:} $\ket{1}\ket{Z}\ket{0}^{\otimes n} $.
\end{algorithm}
The different procedures for dimension reduction using QRDR, QPE-DR, and QPCA are shown in Fig.\cite{fig:flow}.
\begin{figure*}[htbp!]
	\centering
	\includegraphics[width=1\linewidth]{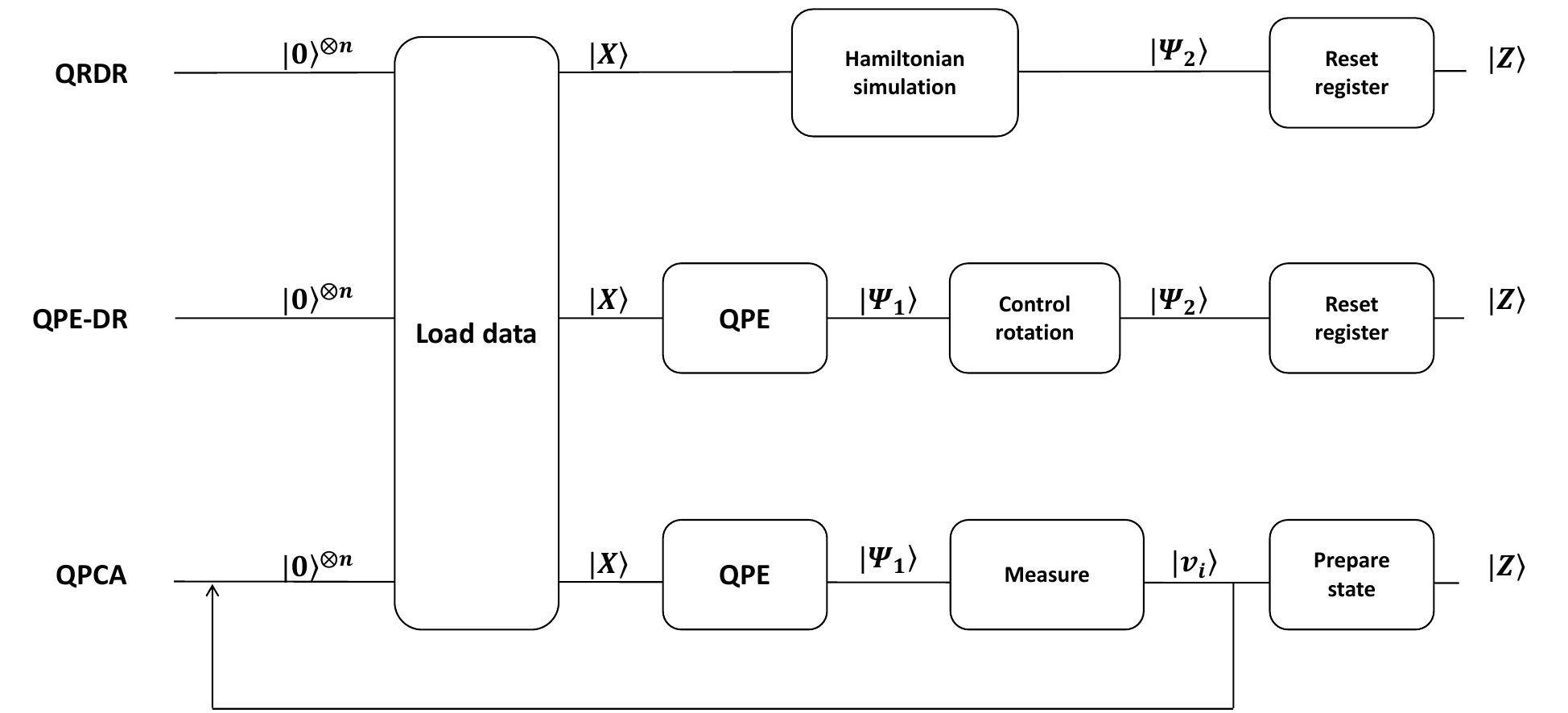}
	\caption{The procedures for dimension reduction using QRDR, QPE-DR, and QPCA differ significantly. The main distinction between QRDR and QPE-DR lies in the use of the Hamiltonian, which replaces the need for QPE and controlled rotation operators in QRDR. In contrast, QPCA requires the use of QPE and involves repeated measurements of auxiliary qubits.}
	\label{fig:flow}
\end{figure*}

\subsection{Errors and Complexity of QRDR}
In this section, we give the details of  the key step of QRDR,  implementing time evolution operator $ e^{-i\mathcal{H}t}$ \cite{Yu.2019,Li_2021}. We provide the specific procedures  that how does the quantum state evolve from Eq. (\ref{Eq:input}) to Eq. (\ref{Eq:evo}). Assuming the input state is the eigenstate of matrix $A$ 
\begin{equation} \label{Eq:eigenstate}
	\ket{\Psi}_{in}=\ket{0}\ket{0}^{\otimes r}\ket{v_k},
\end{equation}
where $k\le R$. Without considering interaction term $\frac{c\pi}{2}\sigma_y\otimes B\otimes I_2^{\otimes n}$, the other parts of Hamiltonian are
\begin{equation}
	\mathcal{H'}=\ket{0}\bra{0}\otimes(\ket{0}\bra{0}^{\otimes r}-I_2^{\otimes r})\otimes I_2^{\otimes n}+\ket{1}\bra{1}\otimes H_p.
\end{equation}
$\ket{\Psi}_{in}$ is the eigenstate of $\mathcal{H'}$ with eigenvalue $0$. For the second part of $\mathcal{H'}$, $\ket{1}\ket{j}\ket{v_k}$ are also the eigenstates with eigenvalues $\lambda_k-\lambda_j$. If $j=k$, it has eigenvalues $\lambda_k-\lambda_j=0$, which are same with $\ket{0}\ket{0}^{\otimes r}\ket{v_k}$.

Firstly, we consider the dynamics between $\ket{0}\ket{0}^{\otimes r}\ket{v_k}$  and $\ket{1}\ket{j}\ket{v_j}$ ($j,k \le R$) because of their same eigenvalues $0$. For each pair of $\ket{0}\ket{0}^{\otimes r}\ket{v_k}$  and $\ket{1}\ket{j}\ket{v_j}$, $\mathcal{H}$ in this subspace can be written as \cite{PhysRevA.93.052334}
\begin{equation}\label{Eq:subspace}
	\mathcal{H}_{kj}=\left[\begin{array}{cc}
		0 & -id_{kj} \\
		id_{kj}^* & 0\\
	\end{array}\right],
\end{equation}
where $d_{kj}=\frac{c\pi}{2}\bra{0}^{\otimes r}B\ket{j}\bra{v_k}I_2^{\otimes n}\ket{v_j}=\frac{1}{2}c\pi\delta_{kj}$. $\delta_{kj}$ is $1$ if and only if $j=k$, and $0$ otherwise. It means that resonant transitions just happens between $\ket{0}\ket{0}^{\otimes r}\ket{v_k}$ and $\ket{1}\ket{k}\ket{v_k}$. So the state in Eq. (\ref{Eq:eigenstate}) after evolution operator $e^{-i\mathcal{H}t}$ is 
\begin{equation}
	e^{-i\mathcal{H}t}\ket{\Psi_{in}}=\cos (\frac{c\pi t}{2})\ket{0}\ket{0}^{\otimes r}\ket{v_k}+\sin(\frac{c\pi t}{2})\ket{1}\ket{k}\ket{v_k}.
\end{equation}
In setting $t=1/c$, the state is $\ket{1}\ket{k}\ket{v_k}$ after this step.

The other case is off-resonance that happens between $\ket{0}\ket{0}^{\otimes r}\ket{v_k}$ and $\ket{1}\ket{j}\ket{v_{k'}}$. Here $j\ne k'$, and in this subspace $\mathcal{H}$ is a little different with Eq. (\ref{Eq:subspace})
\begin{equation}
	\mathcal{H}_{kjk'}=\left[\begin{array}{cc}
		0 & -id_{kjk'} \\
		id_{kjk'}^* & \lambda_{k'}-\lambda_{j}\\
	\end{array}\right].
\end{equation}
Here $d_{kjk'}=\frac{1}{2}c\pi\delta_{kk'}$, so the resonant transition just happens when $ k=k'$ and the amplitude of this small off-resonance is
\begin{equation}
	\begin{aligned}
		b_{kjk'}&=\frac{c\pi |d_{kjk'}|}{\sqrt{(c\pi |d_{kjk'}|)^2+|\lambda_{k'}-\lambda_{j}|^2}}\\
		&=\frac{c\pi\delta_{kk'}}{\sqrt{(c\pi \delta_{kk'})^2+|\Delta_{jk'}|^2}} \\
		&\le \frac{c\pi \delta_{kk'}}{|\Delta_{jk'}|}=\mathcal{O}(c/\Delta_{min}),
	\end{aligned}
\end{equation}
where $\Delta_{min}$ denotes the min erengy gap $\Delta_{\min}= \min {|E_{i+1}-E_i|}$. 
For the other eigenstates that $k>R$, they always fit $k\ne j$ and almost stay in original state with a small off-resonance. 

Therefore, considering the off-resonance error, the input state in Eq. (\ref{Eq:input}) after evolution operator $e^{-i\mathcal{H}/c}$ is
\begin{equation}
	\begin{aligned}
		&\ket{1}\sum_{k=1}^{R}z^k \alpha_k\ket{k}\ket{v_k}
		+\sum_{k'=R+1}^{N}\ket{0}\ket{0}^{\otimes r}z^{k'}\alpha_k'\ket{v_{k'}} \\
		&+\mathcal{O}(c/\Delta_{min})\ket{\Psi}_{e}.
	\end{aligned}
\end{equation}
$\ket{\Psi}_{e}$ is the error state here, and $\alpha_k\ge\sqrt{1-\sum_{j\ne k}b^2_{kjk}}$. Because the sum of $1/n^2$ sequences is converged, it has 
\begin{equation}
	\begin{aligned}
		|\alpha_k|^2&\ge1-\sum_{j\ne k}|b_{kjk}|^2\\
		&\ge 1-(c\pi)^2\sum_{j\ne k}1/|\Delta_{jk}|^2\\
		&\ge 1-(\frac{c\pi}{\Delta_{min}})^2\sum_{j\ne k}1/|j-k|^2\\
		&=1-\mathcal{O}(c^2/\Delta_{min}^2).
	\end{aligned}
\end{equation}
Defining the fidelity as
\begin{equation}\label{eq:error}
	\epsilon=1-|\inner{Z}{\Psi}_{out}|^2 = \mathcal{O}(c^2/\Delta_{min}^2),
\end{equation}
so that the query complexity for time evolution step is 
\begin{equation}
	\begin{aligned}
		&\mathcal{O}(d||\mathcal{H}||_{max}\mathrm{polylog}(MN)t)= \\
		&\mathcal{O}(d\|A\|_{max}\mathrm{polylog}(MN)/c)= \\
		&\mathcal{O}(d\|A\|_{max}\mathrm{polylog}(MN)/\Delta_{min}\epsilon^{0.5}).
	\end{aligned}
\end{equation}
Here $d$ is the sparsity of $A$.

QRDR's time complexity mainly consists of three parts: obtaining eigenvalues, Hamiltonian evolution, and decoupling. Firstly, QRDR requires the eigenvalues of matrix $A$. In previous work \cite{e25010061}, it has been proved that the error $\epsilon$ is proportional to accuracy of eigenvalues, so the complexity of obtaining the first $R$ eigenvalues is $\mathcal{O}(Rd\|A\|_{max}\mathrm{polylog}(MN)/\epsilon)$. For the initial state, it can be prepared in time $\mathcal{O}(\mathrm{poly}\log(MN))$ according to the QRAM \cite{PhysRevLett.100.160501}. In \textit{step 4}, the probability of success is constant if the first $R$ eigenvalues are enough to reconstruct $A$ as $\sum_{k=1}^{R}\sigma_k^2\approx\sum_{k'=1}^{N}\sigma^2_{k'}$.
Secondly, the  query complexity for Hamiltonian evolution is 
$\mathcal{O}(d\|A\|_{max}\mathrm{polylog}(MN)/\Delta_{min}\epsilon^{0.5})$.
Finally, the query complexity of decoupling corresponding to \textit{step 5} is $\mathcal{O}(R^{1.5}d\|A\|_{max}\mathrm{polylog}(MN)/\epsilon)$ \cite{Yu.2019}. We use the method in Ref.\cite{PhysRevLett.118.010501} to achieve the Hamiltonian time evolution operator, and other approaches like VQAs-based methods \cite{yuan2019theory,cirstoiu2020variational,PhysRevX.7.021050} are also appliable.

\begin{figure*}[htbp]
	\centering
	\includegraphics[width=\linewidth]{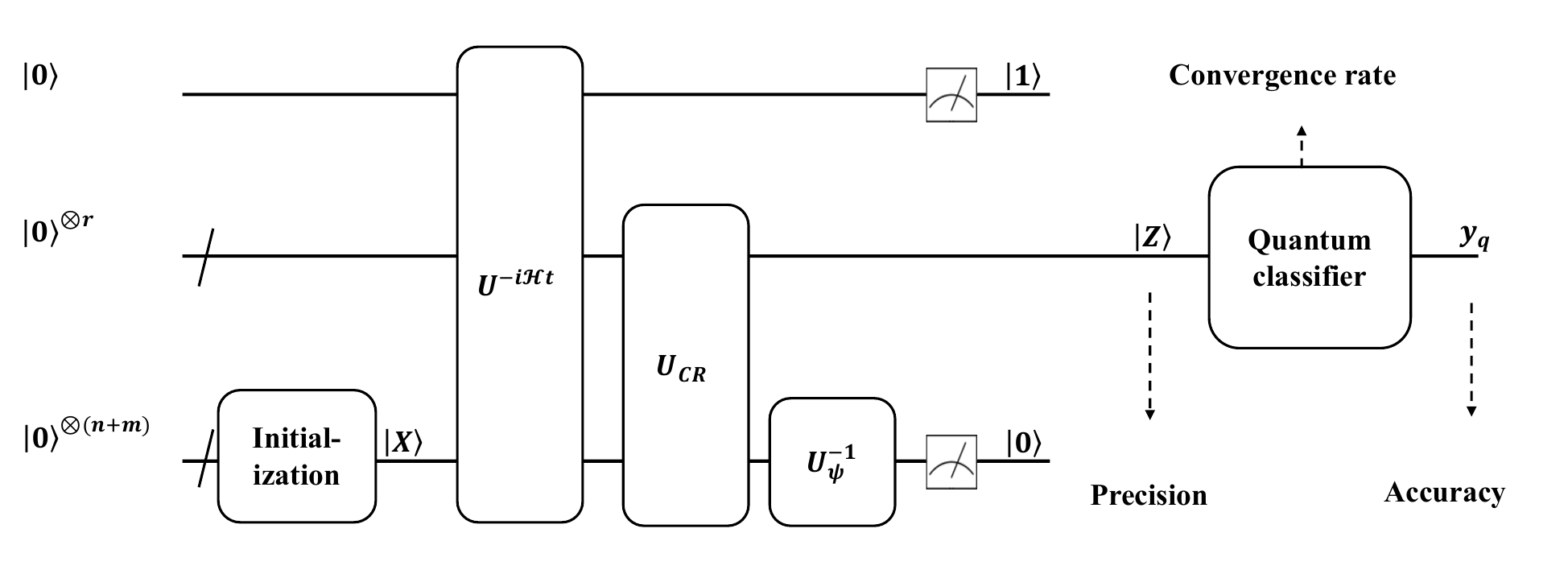}
	\caption{The Schematic quantum circuits of simulating QRDR in quantum machine learning. The two blue parts are the simulated result of quantum algorithms in the classical computer. Three orange dotted line parts are used to calculate precision, convergence rate, and classification accuracy. $|X\rangle$ and $|Z\rangle$ are the quantum states of training data before and after QRDR. Similarly, $|q\rangle$ and $|z_q\rangle $ correspond to the data to be classified. $y_q$ is the predicted label of $y$.}
	\label{fig:simu}
\end{figure*}

In conclusion,  the total time complexity of QRDR is
\begin{equation}
\begin{aligned}
	\mathcal{O}(&Rd\|A\|_{max} \mathrm{polylog }(MN)/\epsilon + \\ 
	&d\|A\|_{max}\mathrm{polylog}(MN)/\Delta_{min}\epsilon^{0.5}+ \\
	&R^{1.5}d\|A\|_{max}\mathrm{polylog}(MN)/\epsilon).
\end{aligned}
\end{equation}
Under the condition that $R=\mathcal{O}(\mathrm{polylog} N)$ eigenvalues can reconstruct the covariance matrix $A$, and $\Delta{\min} = O(\epsilon^{0.5}/R)$, the time complexity is simplified to $\mathcal{O}(d , \mathrm{polylog}(MN) / \epsilon) $.
The following discussion of QPE-based QDR and QPCA assumes the same conditions, where $R=\mathcal{O}(\mathrm{polylog} N)$.
QPE-based QDR \cite{Yu.2019, Li_2021} solve the PCADR problem using quantum phase estimation, which can prepare the eigenvalues in ancilla qubits. As shown in Fig. \ref{fig:flow}, it applies QPE to data state $\ket{x}$ and obtained the state 
\begin{equation}
	\ket{\Psi_1} =\Psi \sum_{i=1}^{M}\sum_{j=0}^{N}z_i^j \ket{\hat{\sigma_j}}\ket{v_j}\ket{i}.
\end{equation}
Then it uses control rotation operator like HHL method to remove the eigenstate corresponding to the eigenvalue after the $R$-th eigenvalue and applies the inverse operation of QPE, then the state is 
\begin{equation}
	\ket{\Psi_2} = \sum_{i=1}^{M}\sum_{j=1}^{R}z_i^j \ket{v_j}\ket{i}.
\end{equation}
$\ket{v_i}$ is the $i$-th eigenstate of $XX^\dagger$.
QRDR can  acheive the same steps  by simulating the Hamiltonian in Eq. \ref{Eq:Hamiltonian}.
The runtime of this QPE-based DR algorithm is $\mathcal{O}(d, \mathrm{polylog}(MN) / \epsilon^3)$. The complexity of QRDR is $\mathcal{O}(d , \mathrm{polylog}(MN) / \epsilon )$ under the assumption that $\|A\|_{\max} $ is $\mathcal{O}(1)$ order. In the circumstances $\Delta{\min} = \mathcal{O}(\epsilon^{0.5}/R)$, our algorithm provides the advantage of polynomial order in accuracy $\epsilon$ compared to QPE-based QDR.
Additionally, QPE requires an extra $\mathcal{O}(1/\epsilon)$ qubits to recover the eigenvalues, whereas QRDR does not require this extra quantum register. QPCA \cite{lloyd2014quantum} and Resonant QPCA \cite{RQPCA_Li} can reduce the dimension of the quantum state, but they must solve each of the $R$ eigenvalues and $M$ samples one by one, as shown in Fig. \ref{fig:flow}, resulting in a complexity of $\mathcal{O}(MR)$. Detailed results are shown in the Table \ref{tb:comp}.

\begin{table*}[htb]
	\caption{Comparison of the complexity of several different quantum dimensionality reduction algorithms with QRDR.}
	\centering
	  \begin{tabular}{|c|c|c|}
	\hline
	& Time complexity & Space complexity \\
	\hline
	QRDR & $O(d, \mathrm{polylog}(MN) / \epsilon)$ & $O(\log R +\log (MN))$ \\
	\hline
	QPE-DR \cite{Li_2021, Yu.2019} & $O(d, \mathrm{polylog}(MN) / \epsilon^3)$ & $O(\log 1/\epsilon +\log R +\log (MN))$ \\
	\hline
	QPCA \cite{RQPCA_Li, lloyd2014quantum} & $O( M\mathrm{polylog}(MN) / \epsilon^3)$ &  $O(\log (MN))$ \\
	\hline
	\end{tabular}
	\label{tb:comp}
\end{table*}

\section{NUMERICAL SIMULATION RESULTS OF QRDR IN QUANTUM MACHINE LEARNING} \label{numer}

In this section, we use the QRDR method to reduce the dimensionality of data prior to its use in quantum machine learning algorithms. First, we assess the precision of our quantum algorithm by comparing the results of quantum dimensionality reduction with theoretical expectations. We then apply two distinct quantum classification algorithms to measure the efficiency of QRDR.

There are two primary reasons for choosing classification as the subsequent task: First, classification tasks allow for easy evaluation of data quality. If the data resulting from QRDR cannot be effectively classified (e.g., through a decrease in final accuracy), it indicates that some useful information may have been lost during the process. Second, accuracy can be used to gauge the convergence rate of quantum machine learning algorithms, which are two key performance indicators for many machine learning tasks. The numerical simulation process is summarized in Fig. \ref{fig:simu}.

\subsection{QRDR FOR QUANTUM SUPPORT VECTORS MACHINE}

In this section, \emph{Connectionist Bench (Sonar, Mines vs. Rocks) Data Set} is used to evaluate our QRDR algorithm. This dataset includes samples of mines and rocks. First, we apply QRDR to reduce the dimensionality of the dataset. Then, we use a quantum support vector machine (QSVM) to classify the data before and after QRDR. The algorithm simulation flow in this section is consistent with that in Fig. \ref{fig:simu}, with taking QSVM as the quantum classfier. The main steps of QRDR were discussed in the previous section, while the steps of QSVM are briefly outlined here.

The input of the quantum support vector machine (QSVM) is a data matrix $A \in \mathbb{R}^{M \times N}$, containing $M$ distinct sample vectors $\bm{x_i}$, each with $N$ parameters. Each $i$-th sample is assigned a label $y_i$ to classify this sample. The objective of QSVM is to train a quantum state in $N + 1$ dimensions, enabling the classification of new vectors $\bm{x'}$ by determining their label $y'$.

The core step of QSVM is to solve the following linear equation using a quantum computer
\begin{equation}\label{inversion}
	\left(\begin{array}{l}
		\eta_0 \\
		\bm{\eta}
	\end{array}\right)=A^{-1}\left(\begin{array}{l}
		0 \\
		\bm{y}
	\end{array}\right), A=\left(\begin{array}{cc}
		0 & \overrightarrow{1}^{T} \\
		\overrightarrow{1} & \mathcal{K}+\gamma^{-1} I
	\end{array}\right).
\end{equation}
$ \mathcal{K}$ is the kernel matrix with the element $\mathcal{K}_{jk}=\bm{x}^T_j\cdot \bm{x}_k$. $\bm{y}=\left(y_{1}, \ldots, y_{M}\right)^{T}$ and $\overrightarrow{1}=(1, \ldots, 1)^{T}$. The hyperplane can be constructed by $\bm{w}=\sum_{j=1}^M\eta_j\bm{x}_j$ and $w_0=\eta_0$. There are several quantum algorithms available for solving this problem, such as the HHL method \cite{PhysRevLett.103.150502}, variational quantum linear solvers, and others \cite{PhysRevA.105.012423, huang2019nearterm}. Once the state $|\eta_0, \bm{\eta}\rangle$ has been prepared, the label of a new sample, $y_i$, can be determined by calculating the inner product of the state $|\eta_0, \bm{\eta}\rangle$ with the new sample state. This paper primarily focuses on dimensionality reduction, so the QSVM component is not covered in detail. For more comprehensive information, please refer to prior literature \cite{Yang_2024, PhysRevA.109.042618, e25010061}.

This dataset contains $M = 208$ samples, each with $N = 60$ features. First, the unitary operator $e^{-i\mathcal{H}t}$ is directly applied to the states. The input state is
\begin{equation}
	\ket{\Psi_{in}}=\sum_{k=1}^{M}\sum_{j=1}^{N}x_k^j\ket{j}\ket{k}.
\end{equation}
\begin{figure}[htbp!]
	\centering
	\includegraphics[width=1\linewidth]{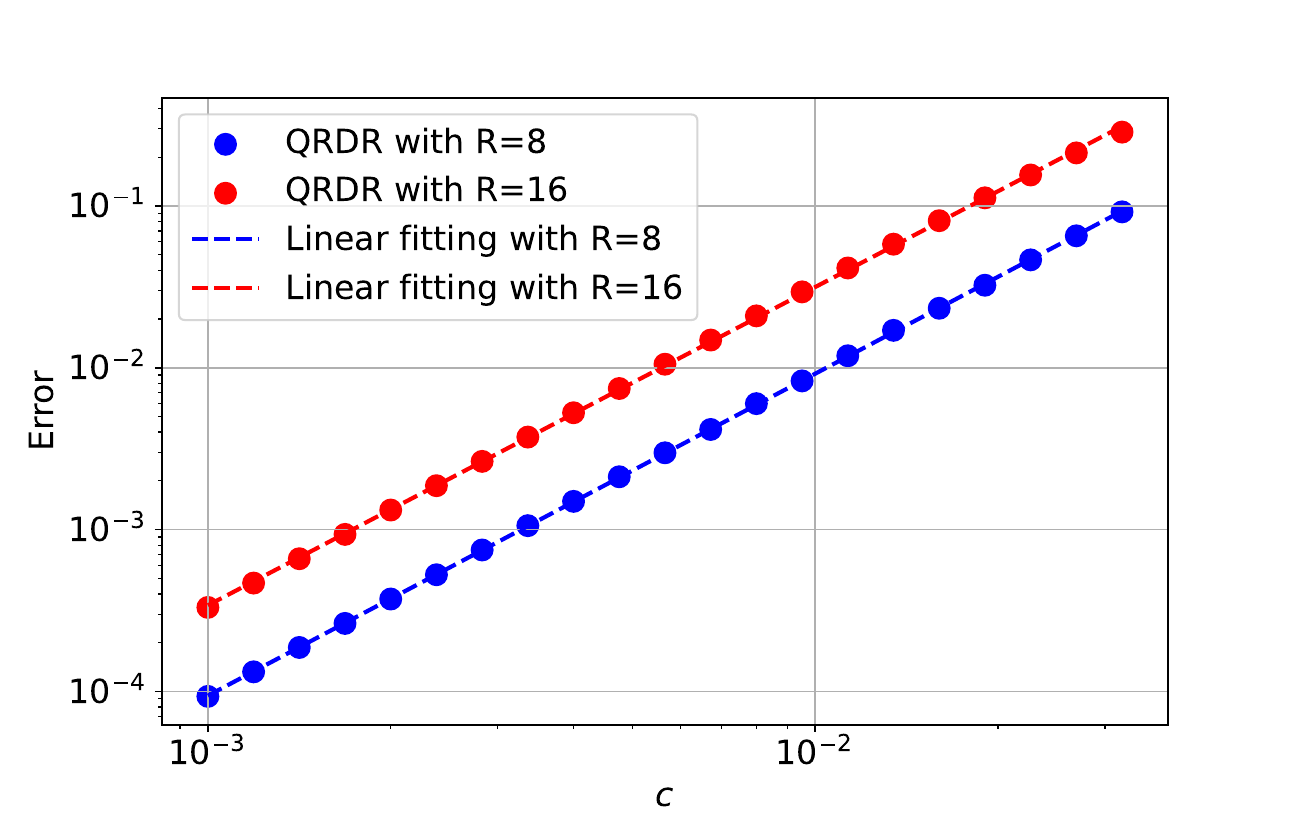}
	\caption{The error of QRDR with different $c$. Red points are simulating results of QSVM in classical computer in $R=16$ and blue line is in $R=8$. The dashed line is the result of a linear fit of error and $c$.}
	\label{fig:pre1}
\end{figure}
We set $R = 16$ and varied $c$ proportionately from $0.001$ to $0.032$. After applying QRDR to the input data, the simulation error is shown in Fig. \ref{fig:simu}. The error defined in Eq. \ref{eq:error} represents the deviation between the numerical simulation results of QRDR and the theoretical results of the classcial DR.

Simulation results indicate that the discrepancy between the QRDR output and theoretical expectations aligns with the theoretical predictions based on the parameter $c$. By calculating the linear fitting coefficients of $\log(1/c)$ and $\log(1/\epsilon^{0.5})$, we can evaluate whether the two parameters are linearly correlated. For $R = 16$, the correlation coefficient is 0.9835, while for $R = 8$, it is 0.9948. This strong linear correlation between $c$ and $\epsilon^{0.5}$ confirms the accuracy of our theoretical analysis. Next, we use the quantum data processed by QRDR and the original quantum data as different input sets for the QSVM. By comparing the training results and complexity of QSVM, we can evaluate the performance of QRDR.

The data is divided into 8 groups, each containing 26 samples, following the standard K-fold cross-validation procedure. Seven sets of data are used as the training set, while one set serves as the test set. Eight QSVM models are trained, and the average accuracy across these models is taken as the final accuracy. For the original data with $N = 60$ features, the accuracy is 0.8625. For $R = 16$, the accuracy improves to 0.8937.

\begin{figure}[htbp!]
	\centering
	\includegraphics[width=1\linewidth]{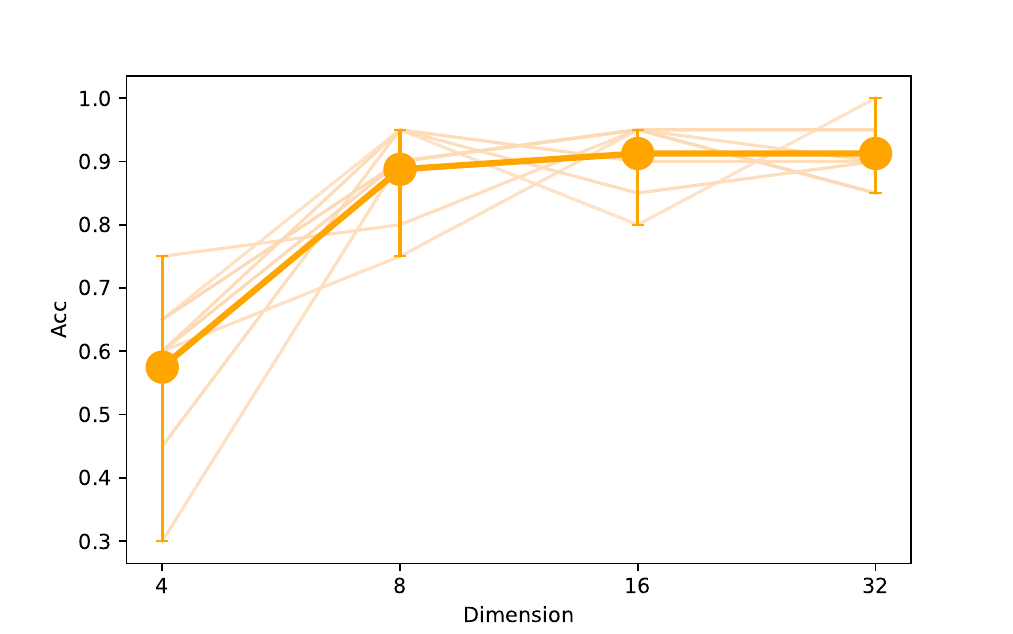}
	\caption{The accuracy of QSVM is presented with varying input data dimensions $R$. The dots represent the average accuracy across multiple calculations, while the upper and lower bounds of the error bars represent the maximum and minimum accuracies observed across multiple calculations. Apart from $R = 4$, QSVM is effective at classifying the data}
	\label{fig:accuracy}
\end{figure}

Next, we select 20 random samples as the validation set and use the remaining data to train the parameters, repeating the process eight times to assess the performance of QSVM with different reduced dimensions $R$ of input data. We observe that QSVM performs very well with $R = 8$. When using higher dimensions like $R = 16$ and $R = 32$, there is only a marginal increase in accuracy. This suggests that only around $8$ parameters are necessary to capture the features of the samples in this dataset. The accuracy of QSVM is illustrated in Fig. \ref{fig:accuracy}.

The complexity of QSVM is $O(\log MN)$, where $N$ is the dimensionality and $M$ is the number of samples \cite{PhysRevLett.113.130503, PhysRevLett.114.140504}. If the dimensionality after QRDR is set to $R = O(\log N)$, the complexity of QSVM with QRDR becomes $O(\log M \log N)$, providing a speedup at this stage.

By simulating the QRDR and QSVM, the error rate of QRDR aligns with our expectations, while the accuracy of QSVM improves and complexity decreases. Our QRDR algorithm enhances the accuracy and speed of QSVM classification, demonstrating its practical value. More details about QSVM, including its workflow and parameter optimization process, can be found in prior work \cite{PhysRevLett.113.130503, Yang_2024}.

\subsection{QRDR FOR QUANTUM MANY BODY PHASE CLASSIFICATION}

In this section, the  quantum phase classification of 1-D Transverse-field Ising model(TFIM) is performed to verify the practicability of our QRDR algorithm.

1D TFIM is a fundamental toy model for quantum phase trasition, which consider the nearest neighbor $ZZ$ interation as well as an extra x-direction tansverse magnetic field in  a spin-half chain, with hamiltonian
\begin{eqnarray}\label{Eq:Ising_hamiltonian}
	H=-J\sum_{i,i+1}\sigma^z_i\sigma^z_{i+1}+h\sum_{i}\sigma^x_i.
\end{eqnarray}

This Hamiltonian is invariant under a global $Z_2$ transformation, $\tau=\prod_i \sigma_x$. Because of the competition of the two non-commute parts in Eq.(\ref{Eq:Ising_hamiltonian}), quantum phase transition appears at the critical point $h/J=1$.  In the region $h>J$, there is a unique ground state that preserve the $Z_2$ symmetry and this system is in a disordered paramagnetic phase. In the region $h<J$, $Z_2$ symmetry is spontaneously broken in the thermodynamic limit, this system is in an ordered ferromagnetic phase.

\begin{figure*}[htbp]
	\label{fig:QCNN_flow}
	\centering
	\includegraphics[width=0.8\linewidth]{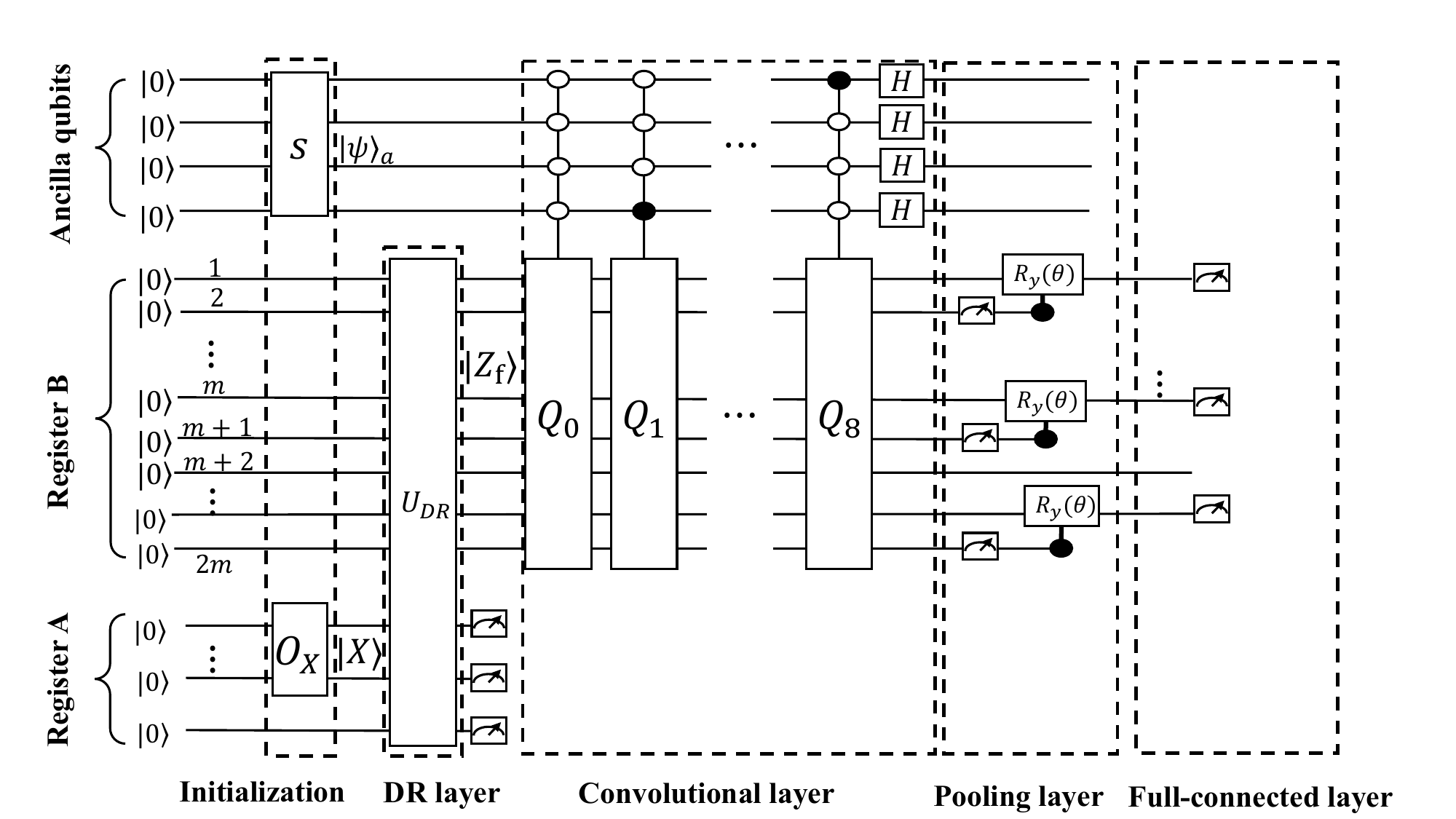}
	\caption{The simplified quantum circuit of QRDR-QCNN algorithm. The initial state $|X\rangle$ is prepared in register A by operator $O_X$, and QRDR utilizes a probe qubit and state $|X\rangle$ to prepare state $|Z\rangle $ after DR in register B. Then, QCNN employs ancillary qubits to process state $|Z\rangle $ with the linear combination of unitary operators $Q_0$, $Q_1$,$\cdots$ $Q_8$ which can act as a  group of convolution operator.}
	
\end{figure*}

We first use QRDR to reduce the dimension of quantum state, and then use a QCNN model based on Ref. \cite{wei2022quantum} to classify the quantum state before and after QRDR respectively. In this case, we simulate a complete quantum convolutional neural network model, including a  dimension reduction layer,  a convolutional layer, a pooling layer, and a full-connected layer, as shown in Fig. \ref{fig:QCNN_flow}.

\begin{figure}
    \centering
      \subfigure[]{\includegraphics[width=1\linewidth,,height=0.1\textwidth]{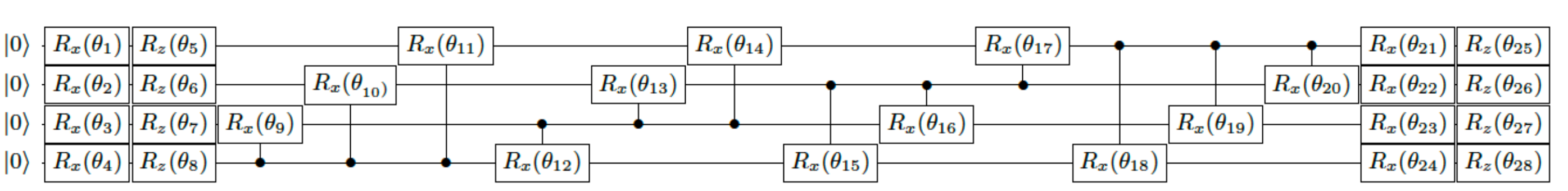}}
      \subfigure[]{\includegraphics[width=0.4\linewidth]{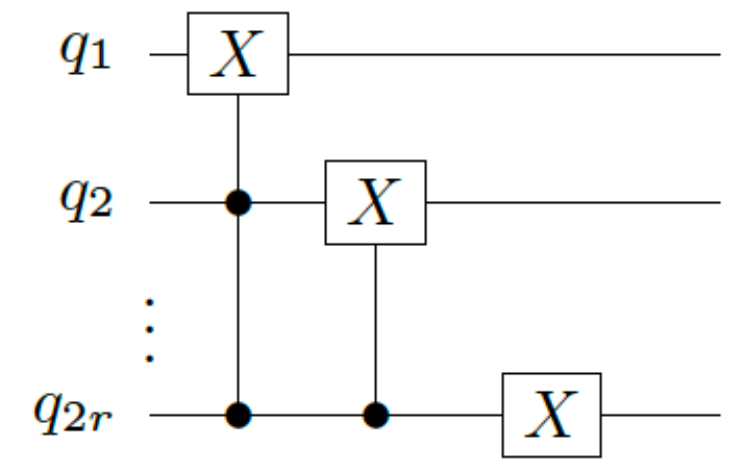}}
       \caption{(a) Quantum circuit for realizing operator $S$. (b) Decomposition of operator $E_1$ in the form of basic gates. }
        \label{S} 
\end{figure}

Operator $S$ is the ansatz to generate superpositon state in the ancillary system. The ansatz\cite{sim2019expressibility} we used for the four ancillary  qubits  contains  28 trainable parameters, and is shown in  Fig. \ref{S}(a). For the convolutional layer,  the nine operators $Q_0$, $Q_1$,$\cdots$, $Q_8$ consist of  the convolution operator. $Q_k$ is the tensor product of two of the following three operators, $E_1$ ,$E_2$ and $E_3$.  $E_2$ is the identity matrix and $\mathbf{E_3}=\mathbf{E_1}^{\dagger}$, where 
\begin{equation}
  E_{1}(i,j)=\begin{cases}
     1 &  (i=j+1),(i=0,j=2^r)  \\
  0 &  Else .\\
\end{cases}\label{eq:U}
\end{equation}

We primarily need to provide the decomposition method for $E_1$. $E_1$ can be constructed using a combination of $O(r^3)$ CNOT gates and Pauli $X$ gates, as illustrated in Fig. \ref{S}(b).

\begin{figure}
	\centering
	\includegraphics[width=1\linewidth]{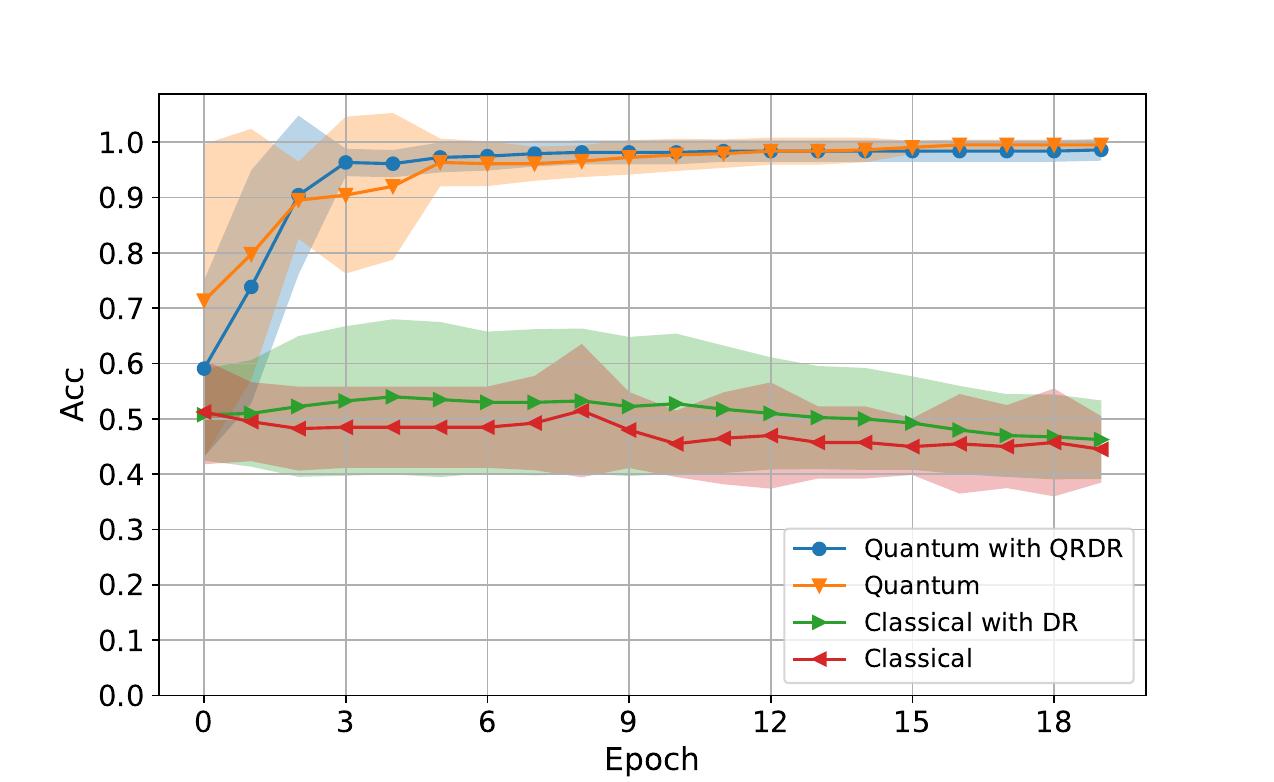}
	\caption{The  performance of QRDR plus QCNN. The blue, orange,  green, and red curves denote the average accuracy of the  QCNN with QRDR, QCNN,  CNN with DR, and CNN, respectively. The shadow areas of the corresponding color denote the accuracy fluctuation range in the 10 times simulation results.}
	\label{fig:qcnn}
\end{figure}

The function of pooling layer after the convolutional layer is to reduce the spatial size  so as to reduce the amount of parameters. Here we choose $R_y(\theta=0)$.   The entire QCNN network consists of 3 layers with a total of 84 trainable parameterized quantum gates. The loss function is related to the expectation value of the parametrized Hamiltonian  $e(p)=\langle{p}| \mathcal{H} \ket{p}$. This Hamiltonian consists of identity operators $I$ and Pauli operators $\sigma_{z}$,
\begin{align}
\mathcal{H}=h^0I+\sum_{i}h^i\sigma_{z}^i+\sum_{i,j}h^{ij}\sigma_{z}^{i}\sigma_{z}^j
\end{align}\label{qubit_hamiltonian}
where $h^0,h^i,h^{ij}$ are the parameters, and Roman indices $i, j$ denote location of the qubit.  
On the other hand, the classical neural network is composed of three fully connected layers, each containing 128 neurons. In both models, we randomly divided a total of 200 data points into 160 training data and 40 testing data. The training set utilized a batch size of 20, and the training process was performed for 20 epochs. The optimizer used was Adam, and the loss function employed was the cross-entropy loss function. 
The numerical simulation is carried out on 'MindSore Quantum'  platform and the  results indicate that QRDR can improve both the performance of  QCNN and CNN. Moreover,  QCNN with QRDR can extremly reduce the training parameter and computation complexity, achieving higher accuracy. The simulation results are shown in Fig. \ref{fig:qcnn}.

The Hamiltonian in Eq. (\ref{Eq:Ising_hamiltonian}) can be directly linked to a p-wave superconducting Hamiltonian via the Jordan-Wigner transformation. In this context, two dangling Majorana fermions can be found at the end of the chain under open boundary conditions. This implies that our QRDR method can be used to distinguish between different symmetry-protected topological phases as well as genuine topological phases.
Topological states, which are independent of local perturbations, have fewer dominant components in the state vector, allowing for natural compression of the state.

\section{DISCUSSION}\label{dis}

To summarize, we propose a novel quantum dimension reduction algorithm called QRDR for preprocessing high-dimensional quantum data prior to the application of quantum machine learning algorithms. 
In comparison to existing quantum DR algorithms, such as QPE-based algorithms and  VQA-based algorithms, it offers distinct advantages in terms of computational complexity and qubit resources by employing QRT. We employ QRDR alongside two different quantum machine learning models, QSVM and QCNN, to evaluate our algorithm's efficacy.  Our method effectively accomplishes two critical tasks: classifying underwater detection targets and analyzing quantum many-body phases.
The simulation results demonstrate its effective reduction of dimensionality in quantum data, thereby enhancing the performance and speed of subsequent algorithms. Given these aforementioned advantages, QRDR is poised to play a significant role in the future era of general quantum computing. Despite our reduction in quantum resource consumption, achieving large-scale algorithm demonstrations on NISQ devices still presents many challenges.


\section{Acknowledgements.} 

S.W. acknowledges the National Natural Science Foundation of China under Grants No. 12005015 and Beijing Nova Program under Grants No. 20230484345; We  acknowledges the National Key Research and Development Program of China (2017YFA0303700); Beijing Advanced Innovation Center for Future Chip (ICFC). P. G. acknowledges the National Natural Science Foundation of China under grant No. 12247164. T.X acknowledges the the National Natural Science Foundation of China No.12275117, Guangdong Basic and Applied Basic Research Foundation No.2022B1515020074. 


\section{Data availability statement} 

The data that support the findings of this study are available on \cite{mq_2021}.



\bibliographystyle{unsrtnat}
\bibliography{QRDR.bib}

\begin{thebibliography}{40}
\providecommand{\natexlab}[1]{#1}
\providecommand{\url}[1]{\texttt{#1}}
\expandafter\ifx\csname urlstyle\endcsname\relax
  \providecommand{\doi}[1]{doi: #1}\else
  \providecommand{\doi}{doi: \begingroup \urlstyle{rm}\Url}\fi

\bibitem[Goodell et~al.(2021)Goodell, Kumar, Lim, and
  Pattnaik]{GOODELL2021100577}
John~W. Goodell, Satish Kumar, Weng~Marc Lim, and Debidutta Pattnaik.
\newblock Artificial intelligence and machine learning in finance: Identifying
  foundations, themes, and research clusters from bibliometric analysis.
\newblock \emph{Journal of Behavioral and Experimental Finance}, 32:\penalty0
  100577, 2021.
\newblock ISSN 2214-6350.
\newblock URL
  \url{https://www.sciencedirect.com/science/article/pii/S2214635021001210}.

\bibitem[Ghoddusi et~al.(2019)Ghoddusi, Creamer, and
  Rafizadeh]{GHODDUSI2019709}
Hamed Ghoddusi, Germán~G. Creamer, and Nima Rafizadeh.
\newblock Machine learning in energy economics and finance: A review.
\newblock \emph{Energy Economics}, 81:\penalty0 709--727, 2019.
\newblock ISSN 0140-9883.
\newblock URL
  \url{https://www.sciencedirect.com/science/article/pii/S0140988319301513}.

\bibitem[Erickson et~al.(2017)Erickson, Korfiatis, Akkus, and
  Kline]{doi:10.1148/rg.2017160130}
Bradley~J. Erickson, Panagiotis Korfiatis, Zeynettin Akkus, and Timothy~L.
  Kline.
\newblock Machine learning for medical imaging.
\newblock \emph{RadioGraphics}, 37\penalty0 (2):\penalty0 505--515, 2017.
\newblock URL \url{https://doi.org/10.1148/rg.2017160130}.
\newblock PMID: 28212054.

\bibitem[Deo(2015)]{doi:10.1161/CIRCULATIONAHA.115.001593}
Rahul~C. Deo.
\newblock Machine learning in medicine.
\newblock \emph{Circulation}, 132\penalty0 (20):\penalty0 1920--1930, 2015.
\newblock URL
  \url{https://www.ahajournals.org/doi/abs/10.1161/CIRCULATIONAHA.115.001593}.

\bibitem[Gujju et~al.(2023)Gujju, Matsuo, and Raymond]{gujju2023quantum}
Yaswitha Gujju, Atsushi Matsuo, and Rudy Raymond.
\newblock Quantum machine learning on near-term quantum devices: Current state
  of supervised and unsupervised techniques for real-world applications, 2023.
\newblock URL \url{https://arxiv.org/abs/2307.00908}.

\bibitem[Dunjko and Briegel(2018)]{Dunjko_2018}
Vedran Dunjko and Hans~J Briegel.
\newblock Machine learning and artificial intelligence in the quantum domain: a
  review of recent progress.
\newblock \emph{Reports on Progress in Physics}, 81\penalty0 (7):\penalty0
  074001, jun 2018.
\newblock URL \url{https://dx.doi.org/10.1088/1361-6633/aab406}.

\bibitem[Kerenidis and Prakash(2016)]{kerenidis2016quantum}
Iordanis Kerenidis and Anupam Prakash.
\newblock Quantum recommendation systems.
\newblock \emph{arXiv preprint arXiv:1603.08675}, 2016.
\newblock URL \url{https://arxiv.org/pdf/1603.08675}.

\bibitem[Rebentrost et~al.(2014)Rebentrost, Mohseni, and
  Lloyd]{PhysRevLett.113.130503}
Patrick Rebentrost, Masoud Mohseni, and Seth Lloyd.
\newblock Quantum support vector machine for big data classification.
\newblock \emph{Physical Review Letters}, 113:\penalty0 130503, Sep 2014.
\newblock URL \url{https://link.aps.org/doi/10.1103/PhysRevLett.113.130503}.

\bibitem[Lloyd et~al.(2014)Lloyd, Mohseni, and Rebentrost]{lloyd2014quantum}
Seth Lloyd, Masoud Mohseni, and Patrick Rebentrost.
\newblock Quantum principal component analysis.
\newblock \emph{Nature Physics}, 10\penalty0 (9):\penalty0 631--633, 2014.
\newblock URL \url{https://www.nature.com/articles/nphys3029}.

\bibitem[Abbas et~al.(2021)Abbas, Sutter, Zoufal, Lucchi, Figalli, and
  Woerner]{abbas2021power}
Amira Abbas, David Sutter, Christa Zoufal, Aur{\'e}lien Lucchi, Alessio
  Figalli, and Stefan Woerner.
\newblock The power of quantum neural networks.
\newblock \emph{Nature Computational Science}, 1\penalty0 (6):\penalty0
  403--409, 2021.
\newblock URL \url{https://doi.org/10.1038/s43588-021-00084-1}.

\bibitem[Beer et~al.(2020)Beer, Bondarenko, Farrelly, Osborne, Salzmann,
  Scheiermann, and Wolf]{beer2020training}
Kerstin Beer, Dmytro Bondarenko, Terry Farrelly, Tobias~J Osborne, Robert
  Salzmann, Daniel Scheiermann, and Ramona Wolf.
\newblock Training deep quantum neural networks.
\newblock \emph{Nature communications}, 11\penalty0 (1):\penalty0 808, 2020.
\newblock URL \url{https://www.nature.com/articles/s41467-020-14454-2}.

\bibitem[Li et~al.(2020)Li, Zhou, Xu, Hu, and Fan]{Li_2021}
YaoChong Li, Ri-Gui Zhou, RuiQing Xu, WenWen Hu, and Ping Fan.
\newblock Quantum algorithm for the nonlinear dimensionality reduction with
  arbitrary kernel.
\newblock \emph{Quantum Science and Technology}, 6\penalty0 (1):\penalty0
  014001, nov 2020.
\newblock URL \url{https://dx.doi.org/10.1088/2058-9565/abbe66}.

\bibitem[Yu et~al.(2019)Yu, Gao, Lin, and Wang]{Yu.2019}
Chao-Hua Yu, Fei Gao, Song Lin, and Jingbo Wang.
\newblock Quantum data compression by principal component analysis.
\newblock \emph{Quantum information processing}, 18\penalty0 (8), 1 2019.
\newblock URL \url{https://doi.org/10.1007/s11128-019-2364-9}.

\bibitem[Liang et~al.(2020)Liang, Shen, Li, and Li]{PhysRevA.101.032323}
Jin-Min Liang, Shu-Qian Shen, Ming Li, and Lei Li.
\newblock Variational quantum algorithms for dimensionality reduction and
  classification.
\newblock \emph{Physical Review A}, 101:\penalty0 032323, Mar 2020.
\newblock URL \url{https://link.aps.org/doi/10.1103/PhysRevA.101.032323}.

\bibitem[Romero et~al.(2017)Romero, Olson, and Aspuru-Guzik]{Romero_2017}
Jonathan Romero, Jonathan~P Olson, and Alan Aspuru-Guzik.
\newblock Quantum autoencoders for efficient compression of quantum data.
\newblock \emph{Quantum Science and Technology}, 2\penalty0 (4):\penalty0
  045001, aug 2017.
\newblock URL \url{https://dx.doi.org/10.1088/2058-9565/aa8072}.

\bibitem[Huang et~al.(2020)Huang, Ma, Yin, Tang, Dong, Chen, Xiang, Li, and
  Guo]{huang2020realization}
Chang-Jiang Huang, Hailan Ma, Qi~Yin, Jun-Feng Tang, Daoyi Dong, Chunlin Chen,
  Guo-Yong Xiang, Chuan-Feng Li, and Guang-Can Guo.
\newblock Realization of a quantum autoencoder for lossless compression of
  quantum data.
\newblock \emph{Physical Review A}, 102:\penalty0 032412, Sep 2020.
\newblock URL \url{https://link.aps.org/doi/10.1103/PhysRevA.102.032412}.

\bibitem[Duan et~al.(2019)Duan, Yuan, Xu, and Li]{PhysRevA.99.032311}
Bojia Duan, Jiabin Yuan, Juan Xu, and Dan Li.
\newblock Quantum algorithm and quantum circuit for a-optimal projection:
  Dimensionality reduction.
\newblock \emph{Physical Review A}, 99:\penalty0 032311, Mar 2019.
\newblock URL \url{https://link.aps.org/doi/10.1103/PhysRevA.99.032311}.

\bibitem[Pan et~al.(2020)Pan, Wan, Liu, Wang, Qin, Wen, and
  Gao]{PhysRevA.102.052402}
Shi-Jie Pan, Lin-Chun Wan, Hai-Ling Liu, Qing-Le Wang, Su-Juan Qin, Qiao-Yan
  Wen, and Fei Gao.
\newblock Improved quantum algorithm for a-optimal projection.
\newblock \emph{Physical Review A}, 102:\penalty0 052402, Nov 2020.
\newblock URL \url{https://link.aps.org/doi/10.1103/PhysRevA.102.052402}.

\bibitem[Yu et~al.(2023)Yu, Lin, and Guo]{YU2023128554}
Kai Yu, Song Lin, and Gong-De Guo.
\newblock Quantum dimensionality reduction by linear discriminant analysis.
\newblock \emph{Physica A: Statistical Mechanics and its Applications},
  614:\penalty0 128554, 2023.
\newblock ISSN 0378-4371.
\newblock URL
  \url{https://www.sciencedirect.com/science/article/pii/S0378437123001097}.

\bibitem[Cong and Duan(2016)]{cong2016quantum}
Iris Cong and Luming Duan.
\newblock Quantum discriminant analysis for dimensionality reduction and
  classification.
\newblock \emph{New Journal of Physics}, 18\penalty0 (7):\penalty0 073011, jul
  2016.
\newblock URL \url{https://dx.doi.org/10.1088/1367-2630/18/7/073011}.

\bibitem[He et~al.(2020)He, Sun, Lyu, and Wang]{he2020quantum}
Xi~He, Li~Sun, Chufan Lyu, and Xiaoting Wang.
\newblock Quantum locally linear embedding for nonlinear dimensionality
  reduction.
\newblock \emph{Quantum Information Processing}, 19:\penalty0 1--21, 2020.
\newblock URL \url{https://doi.org/10.1007/s11128-020-02818-y}.

\bibitem[Sidhu(2019)]{8807145}
Gagan Sidhu.
\newblock Locally linear embedding and fmri feature selection in psychiatric
  classification.
\newblock \emph{IEEE Journal of Translational Engineering in Health and
  Medicine}, 7:\penalty0 1--11, 2019.
\newblock URL \url{https://ieeexplore.ieee.org/document/8807145}.

\bibitem[Li et~al.(2019)Li, Liu, Wang, Ashhab, Cui, Chen, Peng, and
  Du]{PhysRevLett.122.090504}
Zhaokai Li, Xiaomei Liu, Hefeng Wang, Sahel Ashhab, Jiangyu Cui, Hongwei Chen,
  Xinhua Peng, and Jiangfeng Du.
\newblock Quantum simulation of resonant transitions for solving the
  eigenproblem of an effective water hamiltonian.
\newblock \emph{Physical Review Letters}, 122:\penalty0 090504, Mar 2019.
\newblock URL \url{https://link.aps.org/doi/10.1103/PhysRevLett.122.090504}.

\bibitem[Li et~al.(2021)Li, Chai, Guo, Ji, Wang, Shi, Wang, Lloyd, and
  Du]{RQPCA_Li}
Zhaokai Li, Zihua Chai, Yuhang Guo, Wentao Ji, Mengqi Wang, Fazhan Shi,
  Ya~Wang, Seth Lloyd, and Jiangfeng Du.
\newblock Resonant quantum principal component analysis.
\newblock \emph{Science Advances}, 7\penalty0 (34):\penalty0 eabg2589, 2021.
\newblock URL \url{https://www.science.org/doi/abs/10.1126/sciadv.abg2589}.

\bibitem[Giovannetti et~al.(2008)Giovannetti, Lloyd, and
  Maccone]{PhysRevLett.100.160501}
Vittorio Giovannetti, Seth Lloyd, and Lorenzo Maccone.
\newblock Quantum random access memory.
\newblock \emph{Physical Review Letters}, 100:\penalty0 160501, Apr 2008.
\newblock URL \url{https://link.aps.org/doi/10.1103/PhysRevLett.100.160501}.

\bibitem[Wang(2016)]{PhysRevA.93.052334}
Hefeng Wang.
\newblock Quantum algorithm for obtaining the eigenstates of a physical system.
\newblock \emph{Physical Review A}, 93:\penalty0 052334, May 2016.
\newblock URL \url{https://link.aps.org/doi/10.1103/PhysRevA.93.052334}.

\bibitem[Yang et~al.(2023)Yang, Chen, Zhao, Wei, Wen, Wang, Xin, and
  Long]{e25010061}
Fan Yang, Xinyu Chen, Dafa Zhao, Shijie Wei, Jingwei Wen, Hefeng Wang, Tao Xin,
  and Guilu Long.
\newblock Quantum multi-round resonant transition algorithm.
\newblock \emph{Entropy}, 25\penalty0 (1), 2023.
\newblock ISSN 1099-4300.
\newblock URL \url{https://www.mdpi.com/1099-4300/25/1/61}.

\bibitem[Low and Chuang(2017)]{PhysRevLett.118.010501}
Guang~Hao Low and Isaac~L. Chuang.
\newblock Optimal hamiltonian simulation by quantum signal processing.
\newblock \emph{Physical Review Letters}, 118:\penalty0 010501, Jan 2017.
\newblock URL \url{https://link.aps.org/doi/10.1103/PhysRevLett.118.010501}.

\bibitem[Yuan et~al.(2019)Yuan, Endo, Zhao, Li, and Benjamin]{yuan2019theory}
Xiao Yuan, Suguru Endo, Qi~Zhao, Ying Li, and Simon~C Benjamin.
\newblock Theory of variational quantum simulation.
\newblock \emph{Quantum}, 3:\penalty0 191, 2019.
\newblock URL \url{https://quantum-journal.org/papers/q-2019-10-07-191/}.

\bibitem[Cirstoiu et~al.(2020)Cirstoiu, Holmes, Iosue, Cincio, Coles, and
  Sornborger]{cirstoiu2020variational}
Cristina Cirstoiu, Zoe Holmes, Joseph Iosue, Lukasz Cincio, Patrick~J Coles,
  and Andrew Sornborger.
\newblock Variational fast forwarding for quantum simulation beyond the
  coherence time.
\newblock \emph{Npj Quantum Information}, 6\penalty0 (1):\penalty0 1--10, 2020.
\newblock URL \url{https://www.nature.com/articles/s41534-020-00302-0}.

\bibitem[Li and Benjamin(2017)]{PhysRevX.7.021050}
Ying Li and Simon~C. Benjamin.
\newblock Efficient variational quantum simulator incorporating active error
  minimization.
\newblock \emph{Physical Review X}, 7:\penalty0 021050, Jun 2017.
\newblock URL \url{https://link.aps.org/doi/10.1103/PhysRevX.7.021050}.

\bibitem[Harrow et~al.(2009)Harrow, Hassidim, and
  Lloyd]{PhysRevLett.103.150502}
Aram~W. Harrow, Avinatan Hassidim, and Seth Lloyd.
\newblock Quantum algorithm for linear systems of equations.
\newblock \emph{Physical Review Letters}, 103:\penalty0 150502, Oct 2009.
\newblock URL \url{https://link.aps.org/doi/10.1103/PhysRevLett.103.150502}.

\bibitem[Patil et~al.(2022)Patil, Wang, and Krsti\ifmmode~\acute{c}\else
  \'{c}\fi{}]{PhysRevA.105.012423}
Hrushikesh Patil, Yulun Wang, and Predrag~S. Krsti\ifmmode~\acute{c}\else
  \'{c}\fi{}.
\newblock Variational quantum linear solver with a dynamic ansatz.
\newblock \emph{Physical Review A}, 105:\penalty0 012423, Jan 2022.
\newblock URL \url{https://link.aps.org/doi/10.1103/PhysRevA.105.012423}.

\bibitem[Huang et~al.(2019)Huang, Bharti, and Rebentrost]{huang2019nearterm}
Hsin-Yuan Huang, Kishor Bharti, and Patrick Rebentrost.
\newblock Near-term quantum algorithms for linear systems of equations, 2019.
\newblock URL \url{https://arxiv.org/abs/1909.07344}.

\bibitem[Yang et~al.(2024)Yang, Zhao, Wei, Chen, Wei, Wang, Long, and
  Xin]{Yang_2024}
Fan Yang, Dafa Zhao, Chao Wei, Xinyu Chen, Shijie Wei, Hefeng Wang, Guilu Long,
  and Tao Xin.
\newblock A parallel quantum eigensolver for quantum machine learning.
\newblock \emph{New Journal of Physics}, 26\penalty0 (4):\penalty0 043011, apr
  2024.
\newblock URL \url{https://dx.doi.org/10.1088/1367-2630/ad15b4}.

\bibitem[Xu et~al.(2024)Xu, Yang, Wei, Chen, Wei, Wang, Li, and
  Xin]{PhysRevA.109.042618}
Feng Xu, Fan Yang, Chao Wei, Xinyu Chen, Shijie Wei, Hefeng Wang, Jun Li, and
  Tao Xin.
\newblock Quantum simulation of water-molecule bond angles using an nmr quantum
  computer.
\newblock \emph{Physical Review A}, 109:\penalty0 042618, Apr 2024.
\newblock URL \url{https://link.aps.org/doi/10.1103/PhysRevA.109.042618}.

\bibitem[Li et~al.(2015)Li, Liu, Xu, and Du]{PhysRevLett.114.140504}
Zhaokai Li, Xiaomei Liu, Nanyang Xu, and Jiangfeng Du.
\newblock Experimental realization of a quantum support vector machine.
\newblock \emph{Physical Review Letters}, 114:\penalty0 140504, Apr 2015.
\newblock URL \url{https://link.aps.org/doi/10.1103/PhysRevLett.114.140504}.

\bibitem[Wei et~al.(2022)Wei, Chen, Zhou, and Long]{wei2022quantum}
ShiJie Wei, YanHu Chen, ZengRong Zhou, and GuiLu Long.
\newblock A quantum convolutional neural network on nisq devices.
\newblock \emph{AAPPS Bulletin}, 32:\penalty0 1--11, 2022.
\newblock URL \url{https://doi.org/10.1007/s43673-021-00030-3}.

\bibitem[Sim et~al.(2019)Sim, Johnson, and Aspuru-Guzik]{sim2019expressibility}
Sukin Sim, Peter~D Johnson, and Al{\'a}n Aspuru-Guzik.
\newblock Expressibility and entangling capability of parameterized quantum
  circuits for hybrid quantum-classical algorithms.
\newblock \emph{Advanced Quantum Technologies}, 2\penalty0 (12):\penalty0
  1900070, 2019.
\newblock URL \url{https://doi.org/10.1002/qute.201900070}.

\bibitem[Developer(2021)]{mq_2021}
MindQuantum Developer.
\newblock Mindquantum, version 0.6.0, March 2021.
\newblock URL
  \url{https://gitee.com/mindspore/mindquantum/tree/research/paper_with_code/quantum_resonant_dimensionality_reduction_and_application_in_quantum_machine_learning}.

\end{thebibliography}


\end{document}